\documentclass{ws-rv9x6}
\usepackage{ws-rv-van}
\makeindex

\usepackage{graphicx,xcolor,amsmath,amssymb,siunitx,mathtools}
\usepackage[
    colorlinks, citecolor=blue, linkcolor=blue, urlcolor=blue
]{hyperref}

\usepackage{cleveref}
\newcommand{\Refcite}[1]{Ref.~\refcite{#1}}
\newcommand{\Refscite}[1]{Refs.~\citen{#1}}
\newcommand{\Tr}{\operatorname{Tr}}
\newcommand{\U}{\operatorname{U}}
\newcommand{\SU}{\operatorname{SU}}
\newcommand{\A}{\mathrm{A}}
\newcommand{\LL}{\mathrm{L}}
\newcommand{\RR}{\mathrm{R}}
\newcommand{\cc}{\operatorname{c.c.}}

\begin{document}

\begin{center}
    \large\textbf{Light Quarks at Large $N$}
\end{center}

\vspace{1cm}

\author[D. Davies, M. Dine, and B. V. Lehmann]{Daniel Davies,\footnote{dadavies@ucsc.edu} Michael Dine,\footnote{mdine@ucsc.edu} and Benjamin V. Lehmann\footnote{benvlehmann@gmail.com}}

\address{
    Santa Cruz Institute for Particle Physics and Department of Physics,\\University of California, Santa Cruz,\\
    Santa Cruz, CA, USA
}

\begin{abstract}
Lattice gauge theory simulations are our principal probe of the masses of the light quarks. Results from such computations are the primary evidence against the $m_u=0$ solution to the strong CP problem.  The large-$N$ approximation offers an independent approach to light quarks.  We extend existing literature, noting that one can determine the parameters of the non-linear sigma model through second order in quark mass, rule out the $m_u=0$ hypothesis, and make predictions for outputs of lattice calculations and phenomenological fits.  A crucial feature of this analysis is a Wilsonian effective action at scales above the $\eta^\prime$ mass. One can self-consistently test the validity of aspects of this framework, and it may well be good to the part-in-three level. We also note consistency with some phenomenological fits and existing lattice results.
\end{abstract}

\body

\section{Introduction and Overview}

It is a challenge to determine the masses of the light quarks.   Over the past two decades or so, lattice gauge theory has provided measurements of the $u$, $d$, and $s$ quark masses at the few-percent level \cite{pdg}.  One strategy for these analyses is to compute the spectrum of mesons for various choices of quark masses, and to fit the results to determine the parameters of the non-linear sigma model (NLSM). Among the lessons learned, these measurements have ruled out the $m_u=0$ solution of the strong CP problem.  

The large-$N$ approximation, to the extent that it is already reliable for $N=3$, provides an alternative approach to determine the parameters of the NLSM.   Several lattice simulations report that the large-$N$ approximation is already quite good at $N=3$ \cite{latticelargen1,latticelargen2,latticelargen3}, giving some optimism about such a program. At large $N$, at scales above the $\eta^\prime$ mass $m_{\eta^\prime}$, QCD has an approximate $\U(3) \times \U(3)$ symmetry, broken to $\U(3)$.  Below these scales and above the mass scale of the pseudoscalar octet, it has an approximate $\SU(3) \times \SU(3)$ symmetry. As a result, it is natural to consider Wilsonian effective actions in three different energy regimes:  
\begin{enumerate}
    \item  High scales, above the scale of chiral symmetry breaking, $\Lambda_{\chi\mathrm{SB}}$, where the degrees of freedom are quarks and gluons.
    \item  Intermediate scales, below $\Lambda_{\chi\mathrm{SB}}$ and above $m_{\eta^\prime}$, where the degrees of freedom are the octet of pseudoscalar mesons plus the $\eta^\prime$.
    \item  Low scales, below $m_{\eta^\prime}$, where the degrees of freedom are the octet.
\end{enumerate}
Setting aside the $\U(1)$ of baryon number, the symmetries of the first two and third actions differ by the presence of an approximate axial $\U(1)$, i.e., $\U(1)_\A$. Our focus will be on operators quadratic in quark masses in these actions. In the literature, the power counting of $N$ for such operators is often not specific as to which of these actions is under consideration, but it is important to make this division.  In the high-scale action, there are operators allowed by perturbation theory for which the counting of powers of $N$ is straightforward.  These can be matched to operators in the NLSM in the intermediate-scale action.  There are also operators in each of the high- and intermediate-scale actions which violate $\U(1)_\A$.   These effects are inherently non-perturbative.  Instantons would generate such operators in the high-scale action, but these calculations are not under control.  As stressed in \Refcite{wittenlargenchiraldynamics}, in the high-scale action, $\theta$ dependence can be estimated by treating $\theta$ as a spurion, so that, allowing for the anomaly, the $\U(1)_\A$ symmetry is a good symmetry if accompanied by shifts of $\theta$.  Then, considering insertions of $F\tilde F$, each additional power of $\theta$ in the effective action is suppressed by a power of $N$.  Correspondingly, in the matching of the high-scale action to the intermediate-scale action, neglecting quark masses, the latter is a function of $\bigl(\theta - \eta^\prime/f_\pi\bigr) N^{-1}$.

Thus, while instantons are suggestive of violation of the $\U(1)_\A$ symmetry, at large $N$, the behavior with $\theta$ is different than one might naively expect.  This is familiar already, at zeroth order in quark masses, for the question of the $\eta^\prime$ mass.  Instantons suggest a potential for the $\eta^\prime$ which is smooth and periodic in the variable $\bigl(\theta -{\eta^\prime / f_\pi}\bigr)$.  Correspondingly,  instantons predict that in the intermediate-scale action, one should see operators which violate the axial charge by an integer amount.  The leading symmetry-violating effect, however, is a term quadratic in the $\eta^\prime$ field, and the theory has a branched structure \cite{wittenlargenchiraldynamics}; the terms in the effective action do not carry definite $\U(1)_\A$ charge.  When we include quark masses, the masses themselves can be treated as spurions, with definite properties under $\U(1)_\A$.  Quadratic combinations of quark masses possess charges $-2,\dotsc, 2$, predicting a corresponding $e^{iQ\theta}$ dependence on $\theta$ in the action.  Again, because of the $\bigl(\theta - \eta^\prime/f_\pi\bigr) N^{-1}$ dependence of the effective action, terms involving the quark masses must obey the axial symmetry.

As a result of these considerations, the number of terms in the NLSM action at large $N$ is small enough that they can be determined from meson masses and decay constants.   Indeed, in principle, one makes predictions for two decay constants.  With this restriction of the terms in the NLSM action, one can:
\begin{enumerate}
    \item  Compare these results with the phenomenological fits of \Refscite{glstrangequark,bijnens,bijnensupdated}.  These provide a quantitative test of large $N$ for $N=3$, which suggests that the large-$N$ counting for the $\U(1)_\A$-violating operators is correct, and that large $N$ might be good to the part-in-three level or better for $N=3$.
    \item  Compare with lattice results. These similarly provide support for the validity of the large-$N$ picture.
    \item  Self-consistently assess the validity of the large-$N$ approximation for $N=3$.  In particular, loop corrections in the NLSM give access to certain $N$-suppressed contributions to the action, which can be compared to the leading ones.   This comparison again suggests that large-$N$ corrections are $\sim$30\% or smaller.
    \item Make predictions of effects non-linear in light quark masses which should be observable in lattice simulations.  The success of simulations in reproducing these relations can be viewed as a test of the large-$N$ approximation, or as a check on the accuracy of the simulations.
    \item  As noted, rule out the $m_u=0$ solution of the strong CP problem without any input from lattice computations.
 \end{enumerate}
 
In this paper, we review and extend the existing literature on these questions.   We will first recall, in \cref{sec:largen}, some aspects of the NLSM in the large-$N$ approximation.  In particular, we will define our limits of small quark mass and large $N$ as $m_q \propto \epsilon_q / N$, such that at large $N$, the $\eta^\prime$ is light compared to QCD scales, but more massive than the members of the octet of pseudoscalar mesons, consistent with \Refscite{leutwylerlargen,hsuetal}. This will lead us to distinguish Wilsonian effective actions at \textit{three} energy scales, as above. In \cref{sec:instantonoperatorsection}, we detail our argument that at quadratic order in masses, the effective action at large $N$ respects the $\U(1)_\A$ symmetry. In \cref{sec:susy-qcd}, we compare with similar effects in supersymmetric QCD \cite{dinedraperlargen}, which are better theoretically controlled and which support our counting. In \cref{sec:lowestorder}, we determine the NLSM parameters to first order in quark masses and leading order in $N$ (i.e. to order $N^1$).   We then discuss the large-$N$ counting of operators second order in quark masses in \cref{sec:largencounting} for those operators which are generated in perturbation theory.  We do this first at the quark level, considering the problem from a Wilsonian viewpoint.  This Lagrangian can be matched onto the effective Lagrangian for the $\U(3)\times \U(3)$ non-linear sigma model.  We will see that at first order in quark masses and second order in derivatives, the Lagrangian is of order $N$. At second order in quark masses, there are several operators for which the large-$N$ counting can be determined at the perturbative level.

In \cref{sec:spectrumfit}, we consider the derivation of the $\SU(3)\times \SU(3)$ action from the $\U(3)\times \U(3)$ action.  This is simplified by the suppression of effects associated with violation of $\U(1)_\A$.  But we explain, as has been noted in the literature \cite{glstrangequark,bijnens}, that carefully integrating out the $\eta^\prime$ generates an additional operator of order $N$ in the low-energy action.   We recall the formulae for the spectrum and Goldstone boson decay constants in terms of the parameters of the $\SU(3) \times \SU(3)$ NLSM at large $N$, and fit these parameters to measured quantities at leading order in large $N$.  We compare with the results of \Refscite{glstrangequark,bijnens,bijnensupdated}, as well as results from lattice gauge theory summarized in \Refcite{flag}. This provides a measure of the validity of the large-$N$ approximation at $N=3$.

We consider implications of these results in \cref{sec:implications}. We begin by assessing the reliability of the large-$N$ approximation, where we compare large-$N$ results with phenomenological fits in order to assess the size of the $N$-suppressed corrections. We also use the renormalization scale dependence of the NLSM \cite{glstrangequark} to estimate the errors in the leading large-$N$ results.  Both of these tests suggest that the large-$N$ approximation is reliable at the part-in-three level. We further consider the implications of these results for the reliability of lattice simulations themselves, pointing out predicted nonlinear variations in pion mass with quark mass that should be observable in lattice data.  We discuss some results from \Refcite{bmw} from this viewpoint. Finally, we consider the implications for the $m_u=0$ solution of the strong CP problem. We will see that if the instanton operator were of order $N$, the solution would be viable. However, given the absence of the instanton operator and the validity of the perturbative counting, $m_u=0$ can be ruled out. This is consistent with the results of simulations \cite{latticelargen1,latticelargen2,latticelargen3} and with our own estimates for the size of $N$-suppressed corrections. This is also in agreement with statements in the existing literature \cite{leutwylerlargen,Gerard:1989mr}, though with somewhat different reasoning.

In \cref{sec:conclusions}, we summarize, noting that large $N$ seems likely to provide a fair quantitative guide to the features of QCD at $N=3$, consistent with the results of some phenomenological fits and lattice simulations.  To the extent that it is quantitatively valid, the large-$N$ approximation is enough to rule out the $m_u=0$ solution of the strong CP problem, and provides a benchmark to assess the reliability of lattice simulations with very small quark masses.

\section{The Non-Linear Sigma Model at Large  \texorpdfstring{$N$}{N}}
\label{sec:largen}

In this section, we review some features of the chiral lagrangian at large $N$.
As stressed by Witten \cite{wittenlargenchiraldynamics}, the chiral anomaly is suppressed by $N$, and QCD has an approximate $\U(3) \times \U(3)$ symmetry.  Correspondingly, there is a ninth light Goldstone boson, identified with the $\eta^\prime$.  The action is a function of $\theta - \eta^\prime/f_\pi$, and, as argued in \Refcite{wittenlargenchiraldynamics}, each power of $\theta$ adds a power of $N^{-1}$, so only very low orders in $\eta^\prime$ need be considered.  In particular, the $\eta^\prime$ mass-squared is of order $N^{-1}$, while the $\eta^\prime$ interactions are highly suppressed and can be ignored.

As a result, while the degrees of freedom of the NLSM can be described in terms of a $\U(3)$ matrix, it is useful to write this as a phase times an $\SU(3)$ matrix:
\begin{equation}
    \label{eq:U-matrix}
    U =
    \Sigma\exp\left(i \sqrt{\frac23}\frac{\eta^\prime}{2\tilde f_\pi}\right),
    \qquad
    \Sigma = \exp\left(i\frac{\Pi^A \lambda^A}{2 \tilde f_\pi}\right)
    ,
\end{equation}
with $\lambda^A$ the usual Gell-Mann matrices for $A=1,\dotsc,8$, and with $\tilde f_\pi \equiv \frac12\times F_0 \equiv \frac12\times\SI{93}{\mega\electronvolt}$. The leading terms in the NLSM are quite simple:
\begin{equation}
    \label{sec:largenleadingLagrangian}
    \mathcal{L} \supset
    \tilde f_\pi^2 \Tr(\partial_\mu U^\dagger \partial^\mu U) +
    b \Tr(m_q U) + \tfrac{1}{2}m_{\eta^\prime}^2{\eta^\prime}^2
    .
\end{equation}
Here $b$ and $m_q$ depend on renormalization scale and scheme, but the product $bm_q$ does not, so it is convenient to rewrite this, following Gasser and Leutwyler \cite{glstrangequark}, as
\begin{equation}
    \label{eq:largennlsm}
    \mathcal{L} \supset \tfrac14F_0^2\left[
        \Tr(\partial_\mu U^\dagger \partial^\mu U) +
        \Tr(\chi U) + \cc
    \right]
    .
\end{equation}
We will sometimes use an alternative notation:
\begin{equation}
    \label{eq:withtildes}
    \mathcal{L} \supset \tilde f_\pi^2
        \Tr(\partial_\mu U^\dagger \partial^\mu U) +
        2 \tilde f_\pi^2 \Bigl[\Tr(M_q U) + \cc\Bigr],
\end{equation}
where $F_0 = 2 \tilde f_\pi$ and $\chi = 2 M_q$.   We generally prefer the convention of \cref{eq:withtildes} as the form of $U$ in terms of canonical fields is simple.

Our focus is on terms at second order in quark masses.\footnote{Some aspects of the NLSM have been discussed from a holographic point of view in \Refscite{Yadav:2020pmk,Sil:2015xan}. The viewpoint here is
somewhat different and perhaps more conservative.}  More generally, we study operators with two powers of quark mass, or one power of quark mass and two derivatives, or four derivatives.  These are interesting as a potential test of systematic errors in extraction of QCD parameters from lattice computations, and for studying the possibility that $m_u=0$ \cite{georgimcarthur, kaplanmanohar, banksnirseiberg}.  The possible operators in the $\SU(3) \times \SU(3)$ action at this order are enumerated, for example, in \Refscite{glstrangequark,leutwylerlargen,hsuetal}.  One should first consider the operators appearing in the effective action at energy scales large compared to $m_{\eta^\prime}$.  Considering only operators generated in QCD perturbation theory, there are two operators of order $N^1$ (apart from certain four-derivative operators which do not affect the spectrum), as we will review in \cref{sec:largencounting}.  These are:
\begin{equation}
    \label{eq:operators-L5-L8}
    \mathcal L_5 = L_5 \Tr\left(
        \partial_\mu U^\dagger \partial^\mu U \chi U^\dagger + \cc
    \right)
    ,\qquad
    \mathcal L_8 =
    L_8 \Tr\left(\chi^\dagger U \chi^\dagger U + \cc\right)
    .
\end{equation}
Operators of order $N^0$ include:
\begin{equation}
    \label{eq:operators-L6-L7}
    \mathcal L_6 = L_6 \left[
        \Tr\left(\chi^\dagger U + \chi U^\dagger\right)
    \right]^2
    ,\qquad
    \mathcal L_7 = L_7 \left[
        \Tr\left(\chi^\dagger U - \chi U^\dagger\right)
    \right]^2
    .
\end{equation}

\subsection{Non-Perturbative \texorpdfstring{$\U(1)_\A$}{U1A}-Violating Effects}
\label{sec:instantonoperatorsection}

There are phenomena which are inherently non-perturbative in QCD, in the sense that they do not receive \textit{any} contribution in perturbation theory.  For these, the problem of large-$N$ counting is more subtle.  A well-known example is $\theta$-dependence, and, related to this, the potential for the $\eta^\prime$. Witten argued that for $\theta$ one should study $F \tilde F$ correlators at non-zero momentum, and assume that the behavior holds at zero momentum. This leads to the following structure for the potential:
\begin{equation}
    V(\eta^\prime,\theta) =
    N^2\Lambda^4f\left(\frac{\theta - \eta^\prime/f_\pi}{N}\right)
    ,
\end{equation}
and the theory exhibits a branched structure \cite{wittenlargenchiraldynamics}, with no expectation that $f$ is $2\pi$-periodic in $\theta$. This is in contrast to expectations from instantons, where one might expect smooth periodic functions of the form
\begin{equation}
    \sum_n A_n \cos\left[n({\theta - \eta^\prime/f_\pi })\right]
    .
\end{equation}

We can attempt to apply similar reasoning to operators which might appear in an effective action quadratic in quark masses.  We can assign the quarks charge under $\U(1)_\A$. Then the large-$N$ counting above implies that operators of non-zero integer $\U(1)_\A$ are forbidden. Among these disallowed operators is one which we might have expected from instantons.  These are related to the possibility that instantons might generate a substantial contribution to the up quark mass in the case that the tree-level up quark mass were zero \cite{georgimcarthur}.   They would correspond to a term in the $\U(3) \times \U(3)$ Wilsonian action of the form
\begin{equation}
    \label{eq:instantonoperatornlsm}
    \mathcal{L}_{\mathrm{inst}} = L_{\mathrm{inst}}(\det\chi)
        \Tr\left(\chi^{-1} U \right) + \cc
    .
\end{equation}
As we will see, if $L_{\mathrm{inst}}$ were of order $N$, this operator would yield a contribution to an effective $u$ quark mass parameter $(M_u)_{\mathrm{eff}}$ comparable to what is phenomenologically observed.

But this operator is problematic. In the presence of a $\theta$ parameter, treating the quark masses and $\theta$ as spurions, the action is invariant under
\begin{equation}
    q \rightarrow e^{i \alpha} q
    ,\qquad
    m_q \rightarrow e^{-2 i \alpha} m_q
    ,\qquad
    \theta \rightarrow \theta + 6 \alpha
    .
\label{thetatransform}
\end{equation}
Thus,
\begin{equation}
    L_{\mathrm{inst}} \propto e^{i \theta}
    ,
\end{equation}
as would be expected from instantons.  But this is inconsistent with the expectation that each additional factor of $\theta$ is accompanied by a factor of $N^{-1}$. This suggests that there is no local operator of this form in the Wilsonian effective action at scales above $m_{\eta^\prime}$. Loop corrections at lower scales may generate corrections to pseudo-Goldstone masses proportional to $m_d m_s$ in the case that $m_u=0$, for example, but these will be too small to be of interest for the strong CP problem.

\subsection{Aside: Lessons from Supersymmetric QCD}
\label{sec:susy-qcd}
This argument hinges crucially on the counting of zero-momentum $F \tilde F$ insertions, so it is reassuring that this counting holds in a case where one has great control of these effects: supersymmetric QCD at large $N$ with small supersymmetry breaking \cite{dinedraperlargen}.  It is worth recalling how this works, so we can also check effects proportional to powers of quark masses.  In these theories, for zero quark mass there is a non-perturbative
superpotential of the form
\begin{equation}
    W = \Lambda_{\mathrm{hol}}^{3N-N_f}\left(
        \frac{\Phi}{\Lambda_{\mathrm{hol}}}
    \right)^{-\frac{1}{N-N_f}}
    ,\qquad
    \Phi = \det\bar QQ
    ,
\end{equation}
where $N$ is the number of colors and $N_f$ is the number of flavors. The holomorphic scale $\Lambda_{\mathrm{hol}}$ depends on $\theta$ and the conventional $\Lambda$ parameter of QCD as $\Lambda_{\mathrm{hol}} \propto \Lambda\exp\bigl(\frac{i \theta}{3N-N_f}\bigr)$. This form follows from various considerations, but a simple one is that it is the most general consistent with the symmetries of the theory, $\SU(N)_\LL \times \SU(N)_\RR \times \U(1)_R$, where the $\U(1)_R$ is the anomaly-free $R$ symmetry.  It also respects the axial $\U(1)_\A$ symmetry, under which $Q$ and $\bar Q$ rotate by a phase $e^{i\alpha}$, with a compensating shift of the $\theta$ parameter, as in \cref{thetatransform}, or equivalently multiplication of $\Lambda_{\mathrm{hol}}$ by a phase $\exp\bigl(i\frac{2}{3}\frac{\alpha}{N}\bigr)$. Including quark masses (for simplicity taken to be the same), we can compute the holomorphic quantity $\langle W \rangle$, finding
\begin{equation}
    \langle W \rangle \propto\exp\left(i\frac{\theta}{N}\right).
\end{equation}
This holds for all $N>N_f$.  Because it is holomorphic, this is valid for large mass, where the low-energy theory is the pure gauge supersymmetric theory, without matter.  In this case, we have $\langle W \rangle = \langle \lambda \lambda \rangle$ for $\lambda$ the gaugino. If we now include a supersymmetry breaking term $m_\lambda \lambda \lambda$ in the Lagrangian, we obtain a \textit{potential} for $\theta$ exactly as expected from large-$N$ arguments \cite{dinedraperlargen}. A similar result holds for the addition of a term $m_{3/2} W$. More generally, one can add a term
\begin{equation}
    m_{3/2}\int\mathrm{d}^2 \theta\,\mathcal{L}_{\mathrm{eff}} + \cc
    ,
\end{equation}
where we have called the supersymmetry breaking parameter $m_{3/2}$.

Again, this behavior follows purely from symmetry and holomorphy arguments.  We can similarly ask about operators quadratic in quark masses.  Treating the masses as spurions transforming under the flavor symmetries, they transform by a phase $e^{-2i\alpha}$ under $\U(1)_\A$.   This must be compensated by factors of $e^{i \theta} \sim \Lambda_{\mathrm{hol}}^{3N}$. But such terms cannot appear in the effective action, simply by dimensional analysis.  This is at least consistent with the expectations from the arguments above regarding the large-$N$ counting of $\theta$.

\subsection{\texorpdfstring{$\U(3) \times \U(3)$}{U3xU3} vs. \texorpdfstring{$\SU(3) \times \SU(3)$}{SU3xSU3}}
\label{sec:u3su3}

It is important that we have distinguished, here, between the $\U(3) \times \U(3)$ theory and the $\SU(3) \times \SU(3)$ theory at lower scales. In the $\U(3) \times \U(3)$ theory, an operator such as the instanton operator of \cref{eq:instantonoperatornlsm} is indistinguishable from a combination of the operators $\mathcal L_6$, $\mathcal L_7$, and $\mathcal L_8$ in the lower-scale theory.  This follows from an identity for $\SU(3)$ matrices which figures heavily in the observations of Kaplan and Manohar \cite{kaplanmanohar}:
\begin{equation}
    (\det M) \Tr(M^{-1}\Sigma) = \frac12\left\{
        \left[\Tr(M\Sigma^\dagger)\right]^2 -
        \Tr\left(M\Sigma^\dagger M\Sigma^\dagger\right)
    \right\}
    .
    \label{kaplanmanoharidentity}
\end{equation}
If the instanton operator appeared with a coefficient such that $\mathcal{L}_{\mathrm{inst}} \propto N^1$, then this operator \textit{would} be large enough to account for the $u$ quark mass in the case that the ``bare'' mass were zero.  In this case, the coefficients $L_6$, $L_7$, and $L_8$ would be of order $N$, distinct from the counting of quark Feynman diagrams at higher scales.

This counting has a striking implication.  From the spectrum, taking $f_\pi$, $f_K$, $m_\pi^2$, $m_K^2$, $m_\eta^2$ and $m_{\eta^\prime}^2$ from experiment, we can fit the three diagonal entries of $\chi$ (equivalently $M_q$), as well as $L_5$ and $L_8$.  We will do this in \cref{sec:spectrumfit}.

\section{Lowest Order Parameters}
\label{sec:lowestorder}
 
At first order in $M_q$, the masses of the four light mesons are determined by the three parameters $M_u$, $M_d$, and $M_s$ as in \Refcite{weinbergmass}, as well as an electromagnetic contribution $\Delta_\gamma$:
\begin{equation}
    \begin{array}{lll}
          m_{\pi^0}^2 = M_u + M_d,
        & m_{\pi^+}^2 = M_u + M_d + \Delta_\gamma,
        \\[0.25cm]
          m_{K^0}^2 = {M_d + M_s},
        & m_{K^+}^2 = {M_u + M_s} + \Delta_\gamma,
        & m_\eta^2 = \frac13(M_u + M_d + 4 M_s).
    \end{array}
\end{equation}
We can solve for the quark mass parameters in terms of the pion and kaon masses, neglecting the $\eta$ mass, and we obtain the following estimates $\widetilde M_i$ for $M_u$, $M_d$, and $M_s$:
\begin{equation}
    \label{eq:lowestordermasses}
    \widetilde M_u = \SI{6522}{\mega\electronvolt^2}
    ,\quad
    \widetilde M_d = \SI{11698}{\mega\electronvolt^2}
    ,\quad
    \widetilde M_s = \SI{235958}{\mega\electronvolt^2}.
\end{equation}
This corresponds to
\begin{equation}
        \frac{m_d}{m_u} = 1.79,
        \quad
        \frac{m_s}{m_u} = 36.2,
\end{equation}
as in \Refcite{weinbergmass}. At this order, one obtains a prediction $\widetilde m_\eta$ for the $\eta$ mass:
\begin{equation}
    \widetilde m_\eta = 566.3
    .
\end{equation}
The squared mass differs by about 7\% from the observed value, i.e.,
\begin{equation}
    \frac{\widetilde m_\eta^2}{m_\eta^2} = 1.07
    .
\end{equation}
Again, note that in terms of the parameters of \Refcite{glstrangequark}, $\chi = 2 M_q$.

\section{Microscopic Large \texorpdfstring{$N$}{N} Counting}
\label{sec:largencounting}

Having established the suppression of operators quadratic in quark masses which violate the $\U(1)_\A$ symmetry, in this section we discuss large-$N$ counting first in perturbation theory at the quark level, and then its implications for the sigma model effective action.  This counting has been considered in \Refscite{glstrangequark,kaiserleutwuyler,leutwylerlargen}, and we review it here in a slightly different language. We consider terms in a Wilsonian effective action, obtained by integrating out physics above a scale $\Lambda_w$.  Our interest is in terms in this Lagrangian at second order in quark masses.  These include the following operators, written in terms of left-handed two-component fields:
\begin{align}
    &\mathcal{O}_1 = q_{a_f} m_{f \bar f}  \bar q^a_{\bar f}
        q_{b g}^* m_{g\bar g}^* \bar q^{b*}_{\bar g},
    \\
    &\mathcal{O}_2 = q_{a_f} m_{f \bar f}  \bar q^a_{\bar f}
        q_{b g} m_{g\bar g} \bar q^b_{\bar g} + \cc,
    \\
    &\mathcal{O}_3 = q_{a_f} m_{f \bar f}  \bar q^b_{\bar f}
        q_{b g} m_{g\bar g} \bar q^a_{\bar g} + \cc.
\end{align}
Here $a$ and $b$ are color indices, and $f$ and $g$ are flavor indices. These terms are permitted by the symmetries of perturbation theory.  The anomalous term at second order in quark masses generated by one-instanton effects would correspond to:
\begin{equation}
    \mathcal{O}_{\mathrm{SV}} =
    (\det m_q)(m_q^{-1})_{f \bar f} \bar q^a_{\bar f} q_{af }
    .
\end{equation}
We have argued that this operator is forbidden at large $N$.

For $\Lambda_w \sim \Lambda_{\chi\mathrm{SB}}$, at least at the level of $N$ counting, we would expect the microscopic Lagrangian and the $\U(3) \times \U(3)$ NLSM to match. The correspondence between the microscopic operators and the NLSM fields is:
\begin{equation}
    \bar q^a_{\bar f} q_{a f} = b ~U_{\bar f f}
    ,
\end{equation}
where $U$ is as in \cref{eq:U-matrix} and
\begin{equation}
    b= \left|\langle \bar q q \rangle\right| \propto N^1
    .
\end{equation}
Thus the $\mathcal{O}_i$ correspond to the operators $\mathcal{A}_i$ in the sigma model:
\begin{equation}
    \begin{array}{ll}
          \mathcal{A}_1 = \left|{\rm Tr} (m_q U)\right|^2,
        & \mathcal{A}_2 = \bigl(\Tr[m_q U]\bigr)^2,
        \\[0.25cm]
          \mathcal{A}_3 = \Tr(m_q U m_q U),~~~~
        & \mathcal{A}_{\mathrm{SV}}= (\det m_q) \Tr(m_q^{-1} U).
    \end{array}
\end{equation}

Having determined the microscopic operators that appear in the effective action, we now count the powers of $N$ in their coefficients, $\Gamma_i$.  In particular, consider a connected Greens function:
\begin{equation}
    \left\langle
        \bar q(x)^a_{\bar f} q(x)_{a f} \bar q(0)^b_{\bar g} q(0)_{b g}
    \right\rangle
    = A m_{f \bar f} m_{g \bar g} + B m_{f \bar g} m_{g \bar f}.
\end{equation}
Examining connected Feynman diagrams, we find $A \propto N^0$ and $B \propto N^1$.  Comparing this with the insertion of $\Gamma_i$, one can determine the scaling of the couplings $\Gamma_i$ with $N$. Each of these (connected part) is of order $N^2 \Gamma_i$.  Thus, $\Gamma_1$ and $\Gamma_3$ are of order $N^{-2}$, and $\Gamma_2$ is of order $N^{-1}$. This means that $\mathcal{A}_1$ and $\mathcal{A}_2$ appear with coefficients of order $N^0$, while the coefficient of $\mathcal{A}_3$ is of order $N^1$.  In terms of the $L_i$ in \cref{eq:operators-L5-L8,eq:operators-L6-L7}, this yields
\begin{equation}
    L_5 \propto N^1
    ,\quad
    L_6 \propto N^0
    ,\quad
    L_7 \propto N^0
    ,\quad
    L_8 \propto N^1
    .
\end{equation}
Note that one-loop corrections to these leading-order ``tree-level'' results are of the same order in $N$ or suppressed by an additional factor of $N^{-1}$ (in the case of $\mathcal{A}_3$). Because of the absence of non-perturbative $\U(1)_\A$-violating effects, the counting is the same in the $\SU(3) \times \SU(3)$ action, except for $L_7$, which, as we will shortly see, gets a contribution of order $N$ from integrating out the $\eta^\prime$.

\section{Meson Spectrum and Decay Constants at Large \texorpdfstring{$N$}{N}}
\label{sec:spectrumfit}

We turn to effects involving $L_5$, $L_8$, and $\eta$--$\eta^\prime$ mixing, both kinetic and potential, which are of order $M_q^2 N^1$ in the spectrum. We first consider the pseudoscalar meson decay constants.

\subsection{Pseudoscalar Meson Decay Constants}

The coupling $L_5$ modifies the decay constants of the pseudo-Goldstone bosons.  If we keep only the contribution proportional to $m_s$, there is a modification to the $K$ meson and $\eta$ meson decay constants, and a decay constant is generated for the $\eta^\prime$. The coupling $L_5$ modifies the kinetic terms for the mesons, as
\begin{multline}
    \mathcal{L}_{\mathrm{kin}} =
    \tfrac12 \partial_\mu \pi^a \, \partial^\mu \pi^a +
    \tfrac12 \partial_\mu K^A  \, \partial^\mu K^A \left(
        1 + \frac12\frac{L_5 M_s}{\tilde f_\pi^2}
    \right) +
    \tfrac12 \partial_\mu \eta \, \partial^\mu \eta \left(
        1 +  \frac{2}{3}\frac{L_5M_s}{\tilde f_\pi^2}
    \right)
    \\ +
    \tfrac12 \partial_\mu \eta^\prime \, \partial^\mu \eta^\prime \left(
        1 + \frac{1}{3}\frac{L_5 M_s}{\tilde  f_\pi^2}
    \right) +
    \partial_\mu \eta \, \partial^\mu \eta^\prime  \left(
        -\frac{\sqrt{2}}{3}\frac{L_5 M_s}{\tilde f_\pi^2}
    \right)
    .
\end{multline}
We can construct the currents using the Noether procedure, rescaling the meson fields to have canonical kinetic terms. We denote these canonically normalized fields with a hat, e.g. as $\hat\pi^a$. Now we consider transformations of the form
\begin{equation}
    \delta U = \tfrac i2 \lambda^A \omega^A(x) 
    ,
\end{equation}
where $\lambda^A$ are the usual Gell-Mann matrices, so that
\begin{equation}
    \delta \Pi^A = 2\omega^A \tilde f_\pi
    .
\end{equation}
The corresponding currents are
\begin{equation}
    \label{eq:currentsnonlinear}
    \begin{array}{l}
        \displaystyle
        j^a_\mu = f_\pi \partial_\mu \pi^a
            ,\quad
        a=1,\dots,3;
        \\[0.25cm]\displaystyle
        j^A_\mu= f_\pi  \left(1 + \frac{L_5 M_s}{\tilde f_\pi^2}\right)
            \partial_\mu \hat K^A
            ,\quad
        A=1,\dots,4;
        \\[0.4cm]\displaystyle
        j^8_\mu = f_\pi \left(
            1 + \frac43 \frac{L_5 M_s}{\tilde f_\pi^2}
        \right)\partial_\mu \eta -
        \frac{4 \sqrt{2}}{3}\frac{L_5 M_s}{f_\pi^2}\partial_\mu \eta^\prime.
    \end{array}
\end{equation}

The coefficients appearing in the current $j_\mu^8$ can be compared with experimental data. In particular, we define $f_\eta^8$ and $f_{\eta^\prime}^8$ by
\begin{equation}
    j_\mu^8  = f_\eta^8 \partial_\mu \hat \eta +
               f_{\eta^\prime}^8 \partial_\mu \hat \eta^\prime
    ,
\end{equation}
where the hats denote the fields that diagonalize the mass matrix with canonically normalized kinetic terms. We can now extract $f_\eta^8$ and $f_{\eta^\prime}^8$ using the lowest order mixing of $\eta$ and $\eta^\prime$. We first find a numerical value for $L_5$. From experimental data \cite{pdg,pdgsuzuki}, we have $f_K = 1.22 f_\pi$. Thus, setting $M_s = \widetilde M_s$, as in \cref{eq:lowestordermasses}, we obtain
\begin{equation}
    \label{eq:L5-value}
    L_5 = \num{1.8e-3}
    . 
\end{equation}
To proceed, we need the $\eta$--$\eta^\prime$ mass matrix.  The corrections to the eigenvalues of this mass matrix are of order $M_s^2/m_{\eta^\prime}$, i.e. of order $M_s^2 N \sim \epsilon_s^2/N$. In addition to the contributions from \cref{eq:currentsnonlinear}, we should write this in terms of the physical mass eigenstates.  At leading order in the large-$N$ NLSM, the $\eta$--$\eta^\prime$ mass matrix is:
\begin{equation}
    M_{\eta\textnormal{--}\eta^\prime} =
    \begin{pmatrix}
        m_\eta^2 & A \\ A & m_{\eta^\prime}^2
    \end{pmatrix}
    ,\quad
    m_\eta^2 = \tfrac13 \left(M_u + M_d\right) + 4/3 M_s
    ,\quad
    A \simeq -\frac{2\sqrt{2}}{3}M_s.
\end{equation} 
At large $N$, the eigenvalue shifts are of order
\begin{equation}
    \delta m^2 \sim \frac{m_s^2 b^2}{f_\pi^4 m_{\eta^\prime}^2}
               \sim \epsilon_s N^0 m_\eta^2
    .
\end{equation}
Mixings with other singlet states are suppressed by a power of $N$.  At large $N$, then, the lightest physical state is:
\begin{equation}
    \hat\eta = \eta - \frac{A}{m_{\eta^\prime}^2} \eta^\prime
    .
\end{equation}
As a result, we find
\begin{align}
    &f_\eta^8 = \left(1 + \frac{4}{6}\frac{L_5 M_s}{f_\pi^2}\right) f_\pi
    \approx 1.31f_\pi
    ,\\
    &f_{\eta^\prime}^8 =  -\frac{2\sqrt{2}}{6}\frac{L_5 M_s}{f_\pi^2}f_\pi +
        \frac{A}{m_{\eta^\prime}^2}
    \approx -0.48 f_\pi.
\end{align}
In comparison, the measured values of these quantities are given by \Refcite{pdgsuzuki} as
\begin{equation}
    f_\eta^8\approx 1.2 f_\pi
    ,\qquad
    f_{\eta^\prime}^8 \approx -0.45 f_\pi
    .
\end{equation}
These numbers are consistent with errors of order $N^{-1} = 1/3$.

\subsection{Meson Spectrum}

To determine the spectrum at order $N^1 M_q^2$, there are three ingredients:
\begin{enumerate}
\item  The potential for the mesons.
\item  Kinetic mixing, particularly that of the $\eta$ and $\eta^\prime$.
\item  Potential mixing, particularly that of the $\eta$ and $\eta^\prime$.
\end{enumerate}
The potential terms are straightforward.  We first consider kinetic mixing. From the NLSM Lagrangian including $\mathcal L_5$, one has second derivative terms of the form
\begin{equation}
    \mathcal{L}_{\textnormal{kinetic mixing}} =
    \tfrac12(1 + \alpha) (\partial_\mu \eta)^2 +
    \beta \partial_\mu\eta \, \partial^\mu \eta^\prime +
    \tfrac12(1 + \gamma)(\partial_\mu \eta^\prime)^2
    ,
\end{equation}
where we are working to first order in $\alpha$ and $\beta \sim L_5 M_s$. More precisely:
\begin{equation}
    \alpha = \frac{8 L_5 M_s}{3 f_\pi^2}\left(4 M_s + M_u + M_d\right)
    ,\qquad
    \beta = -\frac{2 \sqrt{2}}{3}\left(4 M_s + M_u + M_d\right)
    .
\end{equation}
It is enough to focus on the $\eta$ mass terms, since the leading large-$N$ contribution dominates for the $\eta^\prime$ mass.  Thus, we make the redefinition
\begin{equation}
    \eta = 1 - \tfrac12\alpha\hat\eta - \beta\hat\eta
    .
\end{equation}
Substituting into the potential yields a contribution to the mass involving $\alpha M_s$ and $\beta M_s$. 

The potential mixing of $\eta$ and $\eta^\prime$ is straightforward to evaluate. The relevant terms in the potential are:
\begin{multline}
    \mathcal{L}_{\textnormal{potential mixing}} = 
    \frac12\Biggl[
        \tfrac13\left(M_u + M_d + 4 M_s\right) \eta^2\\
        +\frac{2 \sqrt{2}}{3} \left(M_u + M_d + 4 M_s\right)\eta\eta^\prime +
        m_{\eta^\prime}^2 {\eta^\prime}^2
    \Biggr].
\end{multline}
Putting all of this together yields the following expressions for the $K$ and $\eta$ masses through linear order in $M_s$:
\begin{align}
    & m_{K^0}^2 = M_d + M_s + \frac{4L_8 - 2 L_5}{\tilde f_\pi^2}\left(
        M_s^2 + 2 M_d M_s
    \right)
    ,\label{eq:mesonmasses-k0}
    \\\label{eq:mesonmasses-kpm}
    & m_{K^+}^2 = M_u + M_s + \Delta_\gamma +
        \frac{4L_8 - 2 L_5}{\tilde f_\pi^2}\left(M_s^2 + 2 M_u M_s\right)
    ,
    \\\label{eq:mesonmasses-eta}
    & \begin{multlined}[t]
        m_{\eta}^2 = \tfrac13\left(M_u +M_d +4 M_s\right) +
        \frac{2}{9}\frac{2L_8  - L_5}{\tilde 4 f_\pi^2}\left(
            16 M_s^2 + 8 M_u M_s + 8 M_d M_s
        \right) \\
        + \frac{32}{9}\frac{L_8}{\tilde f_\pi^2} \left[
            M_s^2+ 2 M_s \left(M_u + M_d\right)
        \right]
        -\frac{32}{9}\frac{M_s^2}{m_{\eta^\prime}^2}
        .
    \end{multlined}
\end{align}

\subsection{Determining \texorpdfstring{$L_7$}{L7} and \texorpdfstring{$L_8$}{L8} }

Having now extracted $L_5$ and predicted the meson masses, we turn to the determination of $L_7$ and $L_8$. Again, we assume the validity of large $N$ for $N=3$, and check afterwards whether $1/N$ corrections are small by evaluating one-loop corrections in \cref{sec:largenerrors}.

We can now determine $L_8$ and an associated set of shifts $\delta M_i$ in $M_u$, $M_d$, and $M_s$, i.e., $M_i \to M_i + \delta M_i$. At this order in $M_s$, the pion mass receives no corrections proportional to $L_5$ or $L_8$, which fixes $\delta M_u = -\delta M_d$. By examining our equations for the remaining meson masses, \cref{eq:mesonmasses-k0,eq:mesonmasses-kpm,eq:mesonmasses-eta}, we find
\begin{multline}
    \label{eq:L8-value}
    L_8 = 6.00 \times 10^{-4}
    ,\\
    \delta M_u = \SI{-959}{\mega\electronvolt^2}
    ,\quad
    \delta M_d = \SI{959}{\mega\electronvolt^2}
    ,\quad
    \delta M_s = \SI{47072}{\mega\electronvolt^2}
    .
\end{multline}
We will compare with phenomenological fits \cite{glstrangequark,bijnensupdated} in the next section, where we assess the reliability of the large-$N$ approximation for $N=3$.  But it is interesting to note that the shifts $\delta M_i$ are of order $20\%$, so assuming the validity of large $N$ allows us to quantify the reliability of the expansion in quark mass, and in particular in the strange quark mass.  It would appear that successive orders in the expansion in quark mass (particularly the strange quark mass) are suppressed by factors of order $1/5$.

We can determine the value of $L_7$ after integrating out the $\eta^\prime$, from the requirement that it reproduce the contribution to the $\eta$ mass which arises from $\eta$--$\eta^\prime$ mixing \cite{glstrangequark}.  This yields
\begin{equation}
    L_7 = \frac{2}{3}\frac{f_\pi^2}{m_{\eta^\prime}^2} = \num{1.5e-3}
    ,
\end{equation}
which is about a factor of three larger than the value quoted in \Refscite{glstrangequark,bijnensupdated}.

It is worth noting, as mentioned in \Refcite{glstrangequark} and discussed further in \Refcite{derafael}, that $\mathcal L_7$ is formally of order $N^2$.  Its matrix elements are suppressed by $\epsilon^2/N^2$. The leading order terms are of order $N \epsilon/N$, so there is still a sensible perturbation theory in $N^{-1}$.

\section{Implications for Other Problems}
\label{sec:implications}

Having understood features of the NLSM at large $N$, we now turn to the implications of these observations for significant open problems. We begin in \cref{sec:largenerrors} by examining the reliability of the large-$N$ approximation at $N=3$.  In \cref{phenomenologylessons}, we compare large-$N$ predictions with phenomenological fits and lattice simulations, but we also argue that the size of $N$-suppressed corrections can be estimated self-consistently within the large-$N$ approximation itself. Next, in \cref{sec:implicationsforlattice}, we study implications for lattice simulations, and point out that the large-$N$ approximation makes predictions for the form of the relationship between the pion mass and the quark masses that should be observable in these computations.  Finally, in \cref{sec:implicationsformu=0}, we discuss the consequences of large-$N$ results for the massless-up-quark solution of the strong CP problem.

\subsection{Quantitative Reliability of Large \texorpdfstring{$N$}{N} at \texorpdfstring{$N=3$}{N=3}}
\label{sec:largenerrors}

While the large-$N$ approximation has long been recognized as explaining various qualitative features of the strong interactions, such as Zweig's rule, the existence of narrow resonances, and the like, its quantitative reliability at $N=3$ is less clear.  As we have noted, lattice simulations suggest the agreement may be fairly good for some quantities \cite{latticelargen1,latticelargen2,latticelargen3}. Here, we perform a complementary assessment of the reliability of the large-$N$ approximation in two ways: (1) by comparing large-$N$ predictions with phenomenological fits and lattice simulations, and (2) via the calculation of the leading corrections to the meson masses and decay constants.

We can first ask about the agreement of the large-$N$ limit with phenomenological fits for $N=3$.  In \cref{table:nlsmparameters}, we list several of the $L_i$ values from \Refscite{glstrangequark,bijnensupdated}.  Note that in \Refcite{glstrangequark}, the couplings are renormalized at $m_\eta$, while in \Refcite{bijnensupdated}, they are renormalized at $m_\rho$. Thus, we include in the table the quantity which must be added to the second to compare with the first. The relevant formulae are collected, for example, in \Refcite{glstrangequark}, where the renormalized quantities are labeled as $L_i^r$.  Changing the scale $\mu$ from $\mu = \mu_0 = m_\eta$ to a general $\mu$ is achieved by:
\begin{equation}
    \label{eq:renormalizationformula}
    L_i(\mu) = L_i^r + \frac{\Gamma^i}{16 \pi^2}\log\left(m_\eta/\mu\right)
    ,
\end{equation}
where $\Gamma^5 = 3/8$ and $\Gamma^8 = 5/48$. With these corrections, we see that there is reasonable agreement between the two fits.  This is in agreement with numbers quoted in \Refcite{flag}.

Up to this point, we have determined the values of the NLSM parameters assuming that the large-$N$ approximation is reliable for $N=3$.  To attach uncertainties to these estimates, or to otherwise assess their quantitative validity, we need information about the reliability of the large-$N$ approximation itself.  Without a computation of the corrections to the leading results as a function of $N$, this is not straightforward, and requires assumptions.  We do have some handles, however.

To illustrate the problem, and the potential for large errors, consider the leading results for $L_5$ and $L_8$.  Note that, unlike \Refscite{glstrangequark,bijnensupdated}, we have not specified a renormalization scheme in our computation.  This is because, at leading order in $N$, one-loop effects are suppressed, and there is no scale or scheme dependence to these quantities. We can attempt to address this by including the one-loop corrections, assuming renormalization at a scale $\mu$, and an ultraviolet cutoff scale $\Lambda_{\chi\mathrm{SB}}$. We can then ask: if we add a constant proportional to $N^{-1}$ to achieve agreement with the phenomenological fits, how large is this constant?  If, say, $\Lambda_{\chi\mathrm{SB}} = m_\rho$ and $\mu = m_\eta$, then lower-scale corrections are small, and the shifts in the renormalized parameters are those in the third column of \cref{table:nlsmparameters}. For $L_5$, this correction is about at the $1/3$ level. This is also true for $L_8$. Thus, it is plausible that large $N$ is quantitatively valid at the part-in-three level. This will be our working assumption in the rest of this paper.

\subsection{Lessons from Phenomenological Fits}
\label{phenomenologylessons}

We can ask whether phenomenological fits support the suppression of the instanton operator and the consequent predictions for the $L_i$.  We focus particularly on results of \Refcite{glstrangequark,bijnens}. These fits are to the $\SU(3) \times \SU(3)$ action.  If the instanton operator is present in the $\U(3) \times \U(3)$ action at order $N^1$, then in the $\SU(3) \times \SU(3)$ action, $L_6$ will appear at order $N^1$, as a result of the identity for $\SU(3)$ matrices in \cref{kaplanmanoharidentity}.  In both \Refcite{glstrangequark} and \Refcite{bijnens}, $L_6$ is, in fact, quite small.  This provides some support for the suppression of the instanton operator in the $\U(3) \times \U(3)$ NLSM. The coefficient $L_7$ is in fact somewhat smaller than expected from integrating out the $\eta^\prime$, and $L_4$ is comparable in size to $L_5$.  \Refcite{bijnens} discusses possible explanations for the surprisingly large $L_4$.  If we accept these arguments, and assume that $N$-suppressed corrections to $L_7$ are substantial at $N=3$, the phenomenological fits are compatible with the large-$N$ picture.

\begin{table}[ht]
    \centering
    \tbl{ Phenomenological NLSM parameters. The ``shift'' column indicates the number that must be added to the \Refcite{glstrangequark} values in order to compare with \Refcite{bijnensupdated}, due to the different renormalization scales adopted in these two works.\\}{
    \begin{tabular}{lllll}
    \hline\hline
    \\[-11pt]
    Parameter &  Large $N$ $\times 10^{3}$ &
    \Refcite{glstrangequark} $\times 10^{3}$ &
    \Refcite{bijnensupdated} $\times 10^{3}$ &
    Shift
    \\ [0.5ex]
    \hline
    \\[-11pt]
    $L_4$ & $0 $  &        $  -0.0(5) $ & $\phantom{-}0.3(31) $ & $0.3 $ \\
    $L_5$ & $1.8$  &$\phantom{-}2.2(5) $ & $\phantom{-}1.01(06)$ & $0.8 $ \\
    $L_6$ &  $0$    &      $-0.2(3) $ & $\phantom{-}0.14(85)$ & $0.02$ \\
    $L_7$ &  $1.5 $  &       $ -0.4(15)$ & $          -0.34(08)$ & $0.0$ \\
    $L_8$ & 0.6 & $\phantom{-}1.1(3) $ & $\phantom{-}0.47(07)$ & $0.2 $ \\[0.5ex]
    \hline
    \end{tabular}}
    \label{table:nlsmparameters}
\end{table}

\subsection{Lattice Simulations and the Validity of Large \texorpdfstring{$N$}{N}}
\label{sec:implicationsforlattice}

Lattice simulations provide tests of:
\begin{enumerate}
    \item  Suppression of the $\U(1)_\A$-violating effects.
    \item  Quantitative tests of the predictions of large $N$ for the $L_i$ coefficients.
\end{enumerate}
The FLAG review \cite{flag} reports values of quark masses, the parameter $b$, and the $L_i$ extracted from various simulations.  The $\chi_i$ ($M_i$) extracted from these simulations are within about $10\%$ of the results we have found previously.   This is consistent with the possibility that the large-$N$ corrections at second order in quark mass are of order $30\%$. The $L_i$ are hierarchically ordered, as one might expect at large $N$ with suppressed instanton operator.  In particular, $L_6$ is typically rather small compared to $L_5$ and $L_8$.  Most results are within $30\%$ of the large-$N$ expectation, except for $L_5$, which in some cases is about a factor of $2$ smaller.

\Refcite{bmw} does not quote results for $L_5$ and $L_8$. We have extracted the combination $L_5-2L_8$ by performing a fit to their reported results for the pion mass as a function of $m_{ud} \equiv \frac12(m_u+m_d)$, for a particular choice of lattice spacing, allowing for a constant term, a linear term, and a quadratic term. In principle, $L_5-2L_8$ can be determined by comparing the quadratic and linear coefficients. This is a simplistic procedure, and assumes that poorly-controlled effects contributing to the constant shift do not contaminate the linear or quadratic dependence. However, from this fit, we find that $L_5-2L_8 = \num{9e-5}$, whereas our results from \cref{eq:L5-value,eq:L8-value} imply $L_5-2L_8 = \num{6e-4}$. This is, in some sense, within errors for the large-$N$ result, if we allow 30\% variation of $L_5$ and $L_8$ in either direction, but the result of the fit is still surprisingly small, requiring a near-cancelation between the two coefficients. Current lattice data are not sufficiently precise to test such a cancelation: the results for $L_5-2L_8$ listed in Table 25 of \Refcite{flag} have low precision, and even take different signs in different lattice simulations. Overall, the lattice does not seem to provide a stringent test, at present, of large $N$ in this context. However, as lattice data improve, it will be possible to conduct precise tests of the errors in the large-$N$ approximation, and to directly assess issues such as the presence of such a cancelation between $L_5$ and $L_8$.

\subsection{Massless Up Quark Solution to the Strong CP Problem}
\label{sec:implicationsformu=0}

We now study the implications of the large-$N$ approximation for the massless-up-quark solution of the strong CP problem \cite{georgimcarthur}. We first reframe slightly what one might hope for from this solution. We first consider the situation from the perspective of the $\U(3) \times \U(3)$ action, with no need to include instanton contributions, as discussed in \cref{sec:u3su3}.  Here, there is a $\U(1)_\A$-violating operator,
\begin{equation}
    \mathcal{L}_{\mathrm{inst}} = (\det M_q) \Tr(M_q^{-1} U)
    ,
\end{equation}
which reduces to $(M_u)_{\mathrm{eff}} U_{11}$ as $ M_u \rightarrow 0$.  If $\mathcal{L}_{\mathrm{inst}}$ were of order $N^1$, and equal, say, to $L_8$, the effective $M_u$ would be nearly as large as our fitted $M_u$.  Now we see precisely what would be required at large $N$ to implement this solution: below the scale $m_{\eta^\prime}$, this would require that both $L_6$ and $L_7$ be of order $N^1$. But $L_6$ is of order $N^0$, so one cannot account for the features of the meson spectrum with $M_u=0$. While this point has been made before in the literature \cite{leutwylerlargen}, here we take a slightly different approach, distinguishing between the $\U(3) \times \U(3)$ and $\SU(3)\times\SU(3)$ actions and the corresponding large-$N$ counting.

\section{Conclusions}
\label{sec:conclusions}

It is striking that assuming the validity of the large-$N$ approximation for $N=3$ opens the possibility for quantitative statements about the non-linear sigma model.  It is important, as we have stressed, that one focus carefully on the effective theory at particular scales.  In the Lagrangian at scales above the QCD scale but close to the scale of chiral symmetry breaking, there is a straightforward counting of $N$ for the various possible operators quadratic in quark mass which appear in perturbation theory.  Non-perturbative effects which violate the $\U(1)_\A$ symmetry are highly suppressed at large $N$.  Given the coefficients of the various operators in the high scale action, matching to the NLSM at scales above the $\eta^\prime$ mass is straightforward.

We have recalled that instantons are suggestive of a particular non-perturbative operator in the $\U(3) \times \U(3)$ action which, if its coefficient were of order $N$, would allow the $m_u=0$ solution of the strong CP problem.  But we have seen that if conventional large-$N$ arguments for the dependence of the effective action on $\theta$ and $\eta^\prime$ are correct, such effects are highly suppressed.   We have noted that at strong coupling, such effects can be studied in \textit{supersymmetric} QCD, where the large-$N$ counting indeed holds, and operators quadratic in quark masses which violate anomalous symmetries are highly suppressed with $N$.

Adopting the conventional large-$N$ counting, we have determined the parameters at second order in quark mass, and first order in quark mass and second order in derivatives, in the NLSM.  We have seen that there is rough agreement between these results and each of phenomenological fits and lattice simulations, arguably at the 30\% level. We have seen that from one-loop corrections to the NLSM, one can estimate the size of $N$-suppressed effects to be of order 30\%. This is consistent with expectations from lattice simulations \cite{latticelargen1,latticelargen2,latticelargen3}.

We have discussed three applications of these results. First, we have argued that it is possible to self-consistently estimate the size of higher-order corrections, and thus demonstrate the quantitative reliability of the large-$N$ approximation at $N=3$. A second application is to lattice gauge theory, where the large-$N$ approximation predicts a connection between the low-energy coefficients and the pion mass--quark mass relation, which should be testable with future lattice results. Finally, provided that the large-$N$ approximation is quantitatively reliable, we can rule out the possibility that $m_u=0$, in agreement with \Refscite{kaiserleutwuyler,leutwylerlargen}.

We note that the large-$N$ approximation can in principle serve as a benchmark for lattice results, once sufficient precision is achieved in extracting $L_5-2L_8$. The accuracy of lattice simulations in accounting for effects quadratic in quark masses might have implications for the calculation of the QED vacuum polarization $\Pi(q^2)$, which in turn is relevant to the muon $g-2$ anomaly \cite{bmwg-2}. The status of this anomaly depends on extracting $\Pi(q^2)$ at the 1\% level at momenta of order $m_s \sim m_\mu$. Interestingly, $L_5$ and $L_8$ have the potential to make contributions of this order or larger.

\section*{Acknowledgments}
We thank Tom DeGrand, Patrick Draper and Nathan Seiberg for discussions. This work was supported in part by U.S. Department of Energy grant No. DE-FG02-04ER41286. The work of B.V.L. is supported in part by U.S. Department of Energy grant No. DE-SC0010107 and by the Josephine de Karman Fellowship Trust.

\bibliographystyle{ws-rv-van}
\bibliography{main}

\begin{thebibliography}{24}
\providecommand{\natexlab}[1]{#1}
\providecommand{\url}[1]{\texttt{#1}}
\expandafter\ifx\csname urlstyle\endcsname\relax
  \providecommand{\doi}[1]{doi: #1}\else
  \providecommand{\doi}{doi: \begingroup \urlstyle{rm}\Url}\fi

\bibitem{pdg}
P.~Zyla et~al., {Review of Particle Physics}, \emph{PTEP}. {\bf 2020}\penalty0
  (8), \penalty0 083C01  (2020).
\newblock \doi{10.1093/ptep/ptaa104}.

\bibitem{latticelargen1}
T.~DeGrand and Y.~Liu, {Lattice study of large $N_c$ QCD}, \emph{Phys. Rev. D}.
  {\bf 94}\penalty0 (3), \penalty0 034506  (2016).
\newblock \doi{10.1103/PhysRevD.94.034506}.
\newblock [Erratum: Phys.Rev.D 95, 019902 (2017)].

\bibitem{latticelargen2}
P.~Hern\'andez and F.~Romero-L\'opez, {The large $N_{c}$ limit of QCD on the
  lattice}, \emph{Eur. Phys. J. A}. {\bf 57}\penalty0 (2), \penalty0 52
  (2021).
\newblock \doi{10.1140/epja/s10050-021-00374-2}.

\bibitem{latticelargen3}
G.~S. Bali, L.~Castagnini, B.~Lucini, and M.~Panero, {Large-$N$ mesons},
  \emph{PoS}. {\bf LATTICE2013}, \penalty0 100  (2014).
\newblock \doi{10.22323/1.187.0100}.

\bibitem{wittenlargenchiraldynamics}
E.~Witten, {Large N Chiral Dynamics}, \emph{Annals Phys.} {\bf 128}, \penalty0
  363  (1980).
\newblock \doi{10.1016/0003-4916(80)90325-5}.

\bibitem{glstrangequark}
J.~Gasser and H.~Leutwyler, {Chiral Perturbation Theory: Expansions in the Mass
  of the Strange Quark}, \emph{Nucl. Phys. B}. {\bf 250}, \penalty0 465--516
  (1985).
\newblock \doi{10.1016/0550-3213(85)90492-4}.

\bibitem{bijnens}
J.~Bijnens and I.~Jemos, {A new global fit of the $L^r_i$ at
  next-to-next-to-leading order in Chiral Perturbation Theory}, \emph{Nucl.
  Phys. B}. {\bf 854}, \penalty0 631--665  (2012).
\newblock \doi{10.1016/j.nuclphysb.2011.09.013}.

\bibitem{bijnensupdated}
J.~Bijnens and G.~Ecker, {Mesonic low-energy constants}, \emph{Ann. Rev. Nucl.
  Part. Sci.} {\bf 64}, \penalty0 149--174  (2014).
\newblock \doi{10.1146/annurev-nucl-102313-025528}.

\bibitem{leutwylerlargen}
H.~Leutwyler, {Bounds on the light quark masses}, \emph{Phys. Lett. B}. {\bf
  374}, \penalty0 163--168  (1996).
\newblock \doi{10.1016/0370-2693(96)85876-X}.

\bibitem{hsuetal}
N.~J. Evans, S.~D.~H. Hsu, and M.~Schwetz, {Chiral perturbation theory, large
  N(c) and the eta-prime mass}, \emph{Phys. Lett. B}. {\bf 382}, \penalty0
  138--144  (1996).
\newblock \doi{10.1016/0370-2693(96)00663-6}.

\bibitem{dinedraperlargen}
M.~Dine, P.~Draper, L.~Stephenson-Haskins, and D.~Xu, {$\theta$ and the
  $\eta^\prime$ in Large $N$ Supersymmetric QCD}, \emph{JHEP}. {\bf 05},
  \penalty0 122  (2017).
\newblock \doi{10.1007/JHEP05(2017)122}.

\bibitem{flag}
R.~Sommer, T.~Onogi, and R.~Horsley, {The 2019 lattice FLAG $\alpha_s$
  average}, \emph{PoS}. {\bf ALPHAS2019}, \penalty0 020  (2019).
\newblock \doi{10.22323/1.365.0020}.

\bibitem{bmw}
S.~Durr, Z.~Fodor, C.~Hoelbling, S.~Katz, S.~Krieg, T.~Kurth, L.~Lellouch,
  T.~Lippert, K.~Szabo, and G.~Vulvert, {Lattice QCD at the physical point:
  Simulation and analysis details}, \emph{JHEP}. {\bf 08}, \penalty0 148
  (2011).
\newblock \doi{10.1007/JHEP08(2011)148}.

\bibitem{Gerard:1989mr}
J.~M. Gerard, {The Light Quark Current Mass Ratios and $\eta - \eta^\prime$
  Mixing}, \emph{Mod. Phys. Lett. A}. {\bf 5}, \penalty0 391  (1990).
\newblock \doi{10.1142/S0217732390000457}.

\bibitem{Yadav:2020pmk}
V.~Yadav, G.~Yadav, and A.~Misra, {(Phenomenology/Lattice-Compatible) SU(3)
  M\ensuremath{\chi}PT HD up to $ \mathcal{O} $(p$^{4}$) and the $ \mathcal{O}
  $(R$^{4}$)-Large-N Connection}, \emph{JHEP}. {\bf 08}, \penalty0 151  (2021).
\newblock \doi{10.1007/JHEP08(2021)151}.

\bibitem{Sil:2015xan}
K.~Sil and A.~Misra, {On Aspects of Holographic Thermal QCD at Finite
  Coupling}, \emph{Nucl. Phys. B}. {\bf 910}, \penalty0 754--822  (2016).
\newblock \doi{10.1016/j.nuclphysb.2016.07.014}.

\bibitem{georgimcarthur}
H.~Georgi and I.~N. McArthur, {INSTANTONS AND THE mu QUARK MASS}  (3, 1981).

\bibitem{kaplanmanohar}
D.~B. Kaplan and A.~V. Manohar, {Current Mass Ratios of the Light Quarks},
  \emph{Phys. Rev. Lett.} {\bf 56}, \penalty0 2004  (1986).
\newblock \doi{10.1103/PhysRevLett.56.2004}.

\bibitem{banksnirseiberg}
T.~Banks, Y.~Nir, and N.~Seiberg.
\newblock {Missing (up) mass, accidental anomalous symmetries, and the strong
  CP problem}.
\newblock In \emph{{2nd IFT Workshop on Yukawa Couplings and the Origins of
  Mass}}, pp. 26--41  (2, 1994).

\bibitem{weinbergmass}
S.~Weinberg, {The Problem of Mass}, \emph{Trans. New York Acad. Sci.} {\bf 38},
  \penalty0 185--201  (1977).
\newblock \doi{10.1111/j.2164-0947.1977.tb02958.x}.

\bibitem{kaiserleutwuyler}
R.~Kaiser and H.~Leutwyler, {Large N(c) in chiral perturbation theory},
  \emph{Eur. Phys. J. C}. {\bf 17}, \penalty0 623--649  (2000).
\newblock \doi{10.1007/s100520000499}.

\bibitem{pdgsuzuki}
M.~Suzuki, {Pseudoscalar - Meson Decay Constants: in Review of Particle Physics
  (RPP 1998)}, \emph{Eur. Phys. J. C}. {\bf 3}, \penalty0 353--354  (1998).

\bibitem{derafael}
S.~Peris and E.~de~Rafael, {On the large N(c) behavior of the L(7) coupling in
  chi(PT)}, \emph{Phys. Lett. B}. {\bf 348}, \penalty0 539--542  (1995).
\newblock \doi{10.1016/0370-2693(95)00160-M}.

\bibitem{bmwg-2}
S.~Borsanyi et~al., {Leading hadronic contribution to the muon magnetic moment
  from lattice QCD}, \emph{Nature}. {\bf 593}\penalty0 (7857), \penalty0 51--55
   (2021).
\newblock \doi{10.1038/s41586-021-03418-1}.

\end{thebibliography}

\end{document}